\PassOptionsToPackage{dvipsnames,table,svgname,hyperref}{xcolor}
\documentclass[acmsmall,natbib=true,nonacm,screen]{acmart}
\AtBeginDocument{%
  \providecommand\BibTeX{{%
    \normalfont B\kern-0.5em{\scshape i\kern-0.25em b}\kern-0.8em\TeX}}}
\usepackage{amsmath}
\usepackage{multirow}
\usepackage{bm}
\usepackage{enumitem}
\usepackage{hyperref}
\usepackage{adjustbox}
\usepackage{makecell}
\usepackage{soul}
\usepackage{tikz}
\usepackage{pgfplots}
\usepackage{subcaption}
\usepackage{mathtools}
\usepackage{tabularx}
\usepackage{graphicx}
\usepackage[dvipsnames]{xcolor}
\usetikzlibrary{patterns}
\usepackage{pgfplots}

\definecolor{darkgray176}{RGB}{176,176,176}
\definecolor{lightgray204}{RGB}{204,204,204}
\definecolor{steelblue31119180}{RGB}{31,119,180}

\newtheorem*{assumption}{Assumption}

\newenvironment{customlegend}[1][]{%
\begingroup
\csname pgfplots@init@cleared@structures\endcsname
\pgfplotsset{#1}%
}{%
\csname pgfplots@createlegend\endcsname
\endgroup
}%
\def\addlegendimage{\csname pgfplots@addlegendimage\endcsname}

\usepackage[table]{xcolor}
\usepackage{algorithm}
\usepackage{algpseudocode}

\algnewcommand\algorithmicinput{\textbf{Input:}}
\algnewcommand\Input{\item[\algorithmicinput]}

\algnewcommand\algorithmicoutput{\textbf{Output:}}
\algnewcommand\Output{\item[\algorithmicoutput]}

\definecolor{airforceblue}{rgb}{0.36, 0.54, 0.66}

\usepackage{cleveref}
\definecolor{electricviolet}{rgb}{0.56, 0.0, 1.0}

\author{Daniele Malitesta}
\email{daniele.malitesta@centralesupelec.fr}
\affiliation{\institution{ Université Paris-Saclay, CentraleSupélec, Inria}
    \country{France}
 }

 \author{Emanuele Rossi}
\email{emanuele.rossi1909@gmail.com}
\affiliation{\institution{VantAI}
\country{United States}
 }

 \author{Claudio Pomo}
\email{claudio.pomo@poliba.it}
\affiliation{\institution{Politecnico di Bari}
\country{Italy}
 }

\author{Fragkiskos D. Malliaros}
\email{fragkiskos.malliaros@centralesupelec.fr}
\affiliation{\institution{Université Paris-Saclay, CentraleSupélec, Inria}
    \country{France}
 }

\author{Tommaso Di Noia}
\email{tommaso.dinoia@poliba.it}
\affiliation{\institution{Politecnico di Bari}
\country{Italy}
 }

\renewcommand{\shortauthors}{Malitesta et al.}

\begin{document}

\title[Dealing with Missing Modalities in Multimodal Recommendation]{Dealing with Missing Modalities in Multimodal Recommendation: a Feature Propagation-based Approach}






\begin{abstract}
Multimodal recommender systems work by augmenting the representation of the products in the catalogue through multimodal features extracted from images, textual descriptions, or audio tracks characterising such products. Nevertheless, in real-world applications, only a limited percentage of products come with multimodal content to extract meaningful features from, making it hard to provide accurate recommendations. To the best of our knowledge, very few attention has been put into the problem of missing modalities in multimodal recommendation so far. To this end, our paper comes as a preliminary attempt to formalise and address such an issue. Inspired by the recent advances in graph representation learning, we propose to re-sketch the missing modalities problem as a problem of missing graph node features to apply the state-of-the-art feature propagation algorithm eventually. Technically, we first project the user-item graph into an item-item one based on co-interactions. Then, leveraging the multimodal similarities among co-interacted items, we apply a modified version of the feature propagation technique to impute the missing multimodal features. Adopted as a pre-processing stage for two recent multimodal recommender systems, our simple approach performs better than other shallower solutions on three popular datasets. 
\end{abstract}

\maketitle

\section{Introduction and motivations}
A large percentage of revenues for today's companies come from recommender systems (RSs)~\cite{DBLP:reference/rsh/2011}, machine learning algorithms designed to track preference patterns and provide customers with a personalised surfing experience on online platforms. As with any machine learning model, RSs are eager for useful data sources to derive knowledge about users and items, views/clicks on products, timestamps, co-purchases, or metadata regarding users/items. 

With the advent of the digital area, almost any sort of data can now be uploaded to any website, from images to audio tracks, from texts to videos. Popular online platforms for e-commerce, social media, and multimedia content delivery have begun to host an ever-growing amount of digital data produced by users and/or for the sake of users, such as product images and descriptions, reviews and comments, songs and podcasts, micro-videos and movies. 

Inevitably, customers' experience on online platforms is (also) shaped by such a heterogeneous content they get to interact with. Thus, a user may be driven to buy a specific piece of clothes because she likes its visual patterns or because she agrees with the opinion of another user regarding the same product; another user may greatly enjoy the soundtrack of a movie she is watching, or she could find interesting the lyrics of the song she is listening to.

Hence, it becomes useful (if not imperative) for recommender systems to augment their knowledge regarding users, items, and users' tastes through these \textbf{multimodal} data sources. In specific domains such as fashion~\cite{DBLP:conf/kdd/ChenHXGGSLPZZ19,DBLP:conf/sigir/ChenCXZ0QZ19}, music~\cite{DBLP:conf/sigir/ChengSH16,DBLP:conf/recsys/OramasNSS17,DBLP:conf/bigmm/VaswaniAA21}, food~\cite{DBLP:journals/tmm/MinJJ20,DBLP:journals/eswa/LeiHZSZ21,DBLP:journals/tomccap/WangDJJSN21}, and micro-video~\cite{DBLP:conf/mm/WeiWN0HC19,DBLP:journals/tmm/ChenLXZ21,DBLP:journals/tmm/CaiQFX22} recommendation, multimodal recommender systems (MRSs) have emerged as the leading solutions by overcoming other multimodal-unaware recommendation techniques in terms of accuracy of the generated personalised suggestions. The current literature enumerates a large plethora of proposed multimodal recommendation methods following a multitude of diverse strategies. Nevertheless, the common rationale is to extract high-level features from the multimodal content describing products and use them to power the usual recommendation pipeline. 

Despite their indisputable success, MRSs still face open challenges. For instance, when it comes to real-world settings, a crucial issue is that not all the products in the catalogue may come with multimodal content to extract meaningful multimodal features. In the simplest scenario, the photographs and/or descriptions of some products on an e-commerce platform may be unavailable for any sort of reasons. Indeed, this has the potential to undermine the quality of suggestions provided by the recommender system underneath. Thus, in this work, we tailor the problem of \textbf{missing modalities in multimodal recommendation}. 

To the best of our knowledge, \textbf{very limited (if no) attention} has been put into this aspect so far~\cite{ DBLP:conf/emnlp/WangNL18}. We suppose this literature gap may be ascribed to the ``industrial'' nature of the problem at hand since we are dealing with a scenario/setting which usually applies to real-world applications rather than in vitro experimental setups (as usually done in ``academia''). 

Given the limited literature on the topic, other shortcomings naturally arise. We may summarise them through the following research questions (RQs):
\begin{itemize}[leftmargin=*]
    \item \textbf{RQ1: How is the problem of missing modalities in multimodal recommendation formally defined?} The literature gives no precise definition of the problem, leaving some aspects unclear. For instance, should we define a missing features rate at \textbf{item-} or \textbf{modalities-}level? In other words, is there a subset of items (modalities) missing all (or part of) their features? 
    \item \textbf{RQ2: Are there any standardised datasets to empirically analyse the problem?} From a careful study of the literature, no existing recommendation dataset is tailored to depict the missing modalities problem. In this case, how should we build our own \textbf{realistic} missing modalities datasets to investigate and possibly solve the issue? 
    \item \textbf{RQ3: What are the existing baselines addressing the missing modalities problem?} Apart from traditional approaches commonly exploited in machine learning to pre-process datasets with missing values, to our knowledge, no baseline is currently available regarding the missing modalities problem in multimodal recommendation. How can we test a new approach if no baseline can be compared?
\end{itemize}

\noindent \textbf{Contributions.} Our contributions to the missing modalities problem in multimodal recommendation may be summarised as follows:
\begin{enumerate}[leftmargin=*]
    \item We provide one of the first preliminary problem definitions for missing modalities in multimodal recommendation, that assumes a portion of items in the catalogue does not have any multimodal features. In this respect, we indicate a straightforward method to simulate this setting by starting from popular recommendation datasets and randomly selecting items as the ones with missing modalities. 
    \item Inspired by the latest advances in graph representation learning, we propose to reshape the missing modalities problem to suit the missing graph node features problem. Thus, we can effectively apply the state-of-the-art feature propagation algorithm~\cite{DBLP:conf/log/RossiK0C0B22} (\textbf{FeatProp}) which has recently demonstrated to outperform other similar baselines in imputing missing node features in graphs. Noteworthy, our FeatProp is a \textbf{simple} approach that (by-design) does \textbf{not impact} on the \textbf{training} procedure and can be mounted on top of any multimodal recommender system, making it completely \textbf{model-agnostic}.
    \item To test the effectiveness of FeatProp, we run an \textbf{extensive suite of 1080 experiments} comprising 2 recent multimodal recommender systems (i.e., \texttt{MMSSL}~\cite{DBLP:conf/www/WeiHXZ23}, and \texttt{FREEDOM}~\cite{DBLP:conf/mm/ZhouS23}), 4 methods to impute missing modalities (i.e., Zeros, Mean, Random, FeatProp), 3 recommendation datasets from the Amazon catalogue~\cite{DBLP:conf/sigir/McAuleyTSH15}, 9 percentages of missing modalities, and 5 random samplings. Results largely justify the rationale of our solution. 
\end{enumerate}

The rest of this paper is organised as follows. First, we report on the useful related work and background knowledge regarding the main topics of this work. Second, we elucidate our proposed FeatProp methodology by formalising the missing modalities problem in multimodal recommendation and underlining how we reshape it to suit the graph missing node features problem. Then, we describe the experimental settings we followed to test the efficacy of the FeatProp algorithm, with detailed reproducibility details. Finally, we discuss the obtained results.
\section{Related work}
This section outlines the literature regarding the two main topics addressed in this paper. First, we present the core approaches in multimodal recommendation; then, we describe how previous solutions tackled the problem of missing information in recommendation. 

\subsection{Multimodal recommendation}
In diverse domains such as fashion~\cite{DBLP:conf/kdd/ChenHXGGSLPZZ19,DBLP:conf/sigir/ChenCXZ0QZ19}, music~\cite{DBLP:conf/sigir/ChengSH16,DBLP:conf/recsys/OramasNSS17,DBLP:conf/bigmm/VaswaniAA21}, food~\cite{DBLP:journals/tmm/MinJJ20,DBLP:journals/eswa/LeiHZSZ21,DBLP:journals/tomccap/WangDJJSN21}, and micro-video~\cite{DBLP:conf/mm/WeiWN0HC19,DBLP:journals/tmm/ChenLXZ21,DBLP:journals/tmm/CaiQFX22} recommendation, the inclusion of multimodal content associated with items has proven to significantly enhance the representational capability of recommender systems. With the advancements in multimodal learning~\cite{DBLP:conf/icml/NgiamKKNLN11, DBLP:books/acm/18/BaltrusaitisAM18, DBLP:journals/pami/BaltrusaitisAM19}, multimodal recommender systems (MRSs) strive to address persistent challenges in personalized recommendation, such as data sparsity and cold-start~\cite{DBLP:conf/aaai/HeM16,DBLP:conf/recsys/OramasNSS17,DBLP:conf/mm/VermaGGS20}. Additionally, exploiting multimodal content can help uncover user-item interactions and intentions through attention mechanisms, thereby enhancing the interpretability of recommendations~\cite{DBLP:conf/sigir/ChenZ0NLC17,DBLP:conf/sigir/ChenCXZ0QZ19,DBLP:conf/mm/LiuCSWNK19,DBLP:conf/sigir/Li0YSCZS21}.

Given the recent surge in graph neural networks~\cite{DBLP:journals/tnn/ScarselliGTHM09,DBLP:journals/corr/abs-2104-13478} in recommendation~\cite{DBLP:conf/sigir/0001DWLZ020, DBLP:conf/cikm/MaoZXLWH21, DBLP:conf/sigir/PengSM22, DBLP:conf/iclr/Cai0XR23}, several techniques have begun integrating multimodality into user-item bipartite graphs and knowledge graphs~\cite{DBLP:journals/ipm/TaoWWHHC20, DBLP:conf/cikm/SunCZWZZWZ20,  DBLP:journals/tmm/WangWYWSN23}. These techniques refine the multimodal representations of users and items through various approaches employing the message-passing schema. While initial efforts involved simply injecting multimodal item features into the graph-based pipeline~\cite{DBLP:conf/kdd/YingHCEHL18, DBLP:conf/mm/WeiWN0HC19}, more advanced methods now learn separate graph representations for each modality and disentangle users' preferences at the modality level~\cite{DBLP:journals/ipm/TaoWWHHC20,DBLP:conf/mm/WeiWN0C20,DBLP:conf/cikm/KimLSK22}. Recent approaches also concentrate on unveiling multimodal structural distinctions among items in the catalogue~\cite{DBLP:conf/mm/Zhang00WWW21,DBLP:conf/mm/LiuYLWTZSM21,DBLP:conf/mir/LiuMSO022, DBLP:conf/mm/ZhouS23}, sometimes leveraging self-supervised~\cite{DBLP:conf/www/ZhouZLZMWYJ23, DBLP:conf/www/WeiHXZ23} and contrastive~\cite{DBLP:conf/sigir/Yi0OM22} learning.

In this work, we propose FeatProp, a solution for missing modalities in multimodal recommendation that can be placed as a \textbf{pre-processing} module on top of any multimodal recommender system. Specifically, we assess its goodness when jointly applied to two recent approaches, namely, \texttt{MMSSL}~\cite{DBLP:conf/www/WeiHXZ23}, and \texttt{FREEDOM}~\cite{DBLP:conf/mm/ZhouS23}.

\subsection{Missing information in recommendation}
In a broader sense, the issue of missing information has always been under debate in recommendation. The literature outlines two main research directions regarding missing information on (i) user-item feedback and (ii) content/metadata.

Indubitably, the \textbf{missing feedback} issue is widely recognised as a crucial challenge. On the one hand~\cite{DBLP:journals/siamrev/Strawderman89, 10.1093/biomet/63.3.581}, the underneath assumption is that ratings in real-world recommendation data are \textbf{missing at random} since the missing probability does not depend on the ratings themselves; in such a context, learning from observed (not-missing) data is sufficient to provide optimal predictions. On the other hand~\cite{DBLP:journals/siamrev/Strawderman89, 10.1093/biomet/63.3.581}, a more complex and interesting scenario is the one of \textbf{missing not at random} ratings, where multiple mechanisms determine the missing feedback ratings; hence, the recommendation algorithm should not ignore these mechanisms during the training. Indeed, the latter setting has been largely investigated over the last few decades~\cite{DBLP:conf/kdd/Steck10, DBLP:conf/ijcai/MarlinZRS11,  DBLP:conf/recsys/LimML15, DBLP:conf/recsys/YangCXWBE18, DBLP:conf/nips/WangGZZ18, DBLP:journals/tkde/ZhengWXLW22, DBLP:conf/icml/WangZ0Q19, DBLP:conf/wsdm/SaitoYNSN20, DBLP:conf/iclr/LiZ023}, sometimes tailoring the issue to certain domains and tasks such as multicriteria~\cite{DBLP:conf/nss/Takasu11}, social~\cite{DBLP:conf/icdm/Chen0ESFC18}, and movies recommendation~\cite{DBLP:conf/dsaa/VernadeC15}.

Conversely, the issue of \textbf{missing content} in recommendation has been poorly explored in the related literature so far. Based on a careful review, the two principal application scenarios involve either items' \textbf{metadata}~\cite{DBLP:conf/cikm/ShiZYZHLM19, DBLP:journals/tkde/LiuCZLN22} or \textbf{multimodal} content describing products~\cite{DBLP:conf/emnlp/WangNL18}. On the one hand, the works in~\cite{DBLP:conf/cikm/ShiZYZHLM19, DBLP:journals/tkde/LiuCZLN22} address the problem of missing attribute entries in the users' and items' metadata following two different strategies, namely, feature sampling towards model's robustness~\cite{DBLP:conf/cikm/ShiZYZHLM19} and graph convolutional networks propagating features on a tripartite graph of users, items, and attributes nodes~\cite{DBLP:journals/tkde/LiuCZLN22}. On the other hand, the solution presented in~\cite{DBLP:conf/emnlp/WangNL18} tackles the missing modalities issue in multimodal recommendation through a generative technique which is trained to reconstruct specific missing modalities embeddings; the only modalities involved are visual and textual.

Indeed, our work settles in the same research direction as the \textbf{missing content} one. As highlighted, the literature regarding this issue is quite limited, especially if we focus on multimodal recommendation (i.e., the only related work is~\cite{DBLP:conf/emnlp/WangNL18}). Differently from the existing methods, FeatProp stands out as a \textbf{pre-processing module} that can be placed on top of any multimodal recommender system, is a \textbf{simple} approach that (by-design) does not impact on the overall \textbf{training} process, and is a \textbf{generalised solution} accounting for visual, textual, audio, and \textbf{all sort of possible modalities}.
\section{Background}
This section provides useful notions and formulas regarding the technologies adopted for this paper. After having introduced and defined the multimodal recommendation task, we describe the issue of missing node features in graph representation learning, and eventually summarise the core concepts underlying the original FeatProp algorithm~\cite{DBLP:conf/log/RossiK0C0B22} used to address such an issue. 

\subsection{Multimodal recommendation}
In a recommendation system, let $\mathcal{U}$ and $\mathcal{I}$ be the sets of users and items, respectively. Then, assuming an implicit feedback scenario, we indicate with $\mathbf{R}^{|\mathcal{U}| \times |\mathcal{I}|}$ the user-item interaction matrix, where $\mathbf{R}_{ui} = 1$ if there exists a recorded interaction between user $u \in \mathcal{U}$ and item $i \in \mathcal{I}$, 0 otherwise.

As in any latent factor-based approaches for recommendation (e.g., matrix factorisation~\cite{DBLP:journals/computer/KorenBV09}), we map users' and items' IDs to embeddings in the latent space, where $\mathbf{E}_u \in \mathbb{R}^k$ and $\mathbf{E}_i \in \mathbb{R}^k$ represent the $k$-dimensional embeddings for user $u \in \mathcal{U}$ and item $i \in \mathcal{I}$, with $k \ll |\mathcal{U}|, |\mathcal{I}|$. On such a basis, the recommendation task is about reconstructing the user-item interaction matrix $\hat{\mathbf{R}}$ by learning meaningful representations for the user and item embeddings $\mathbf{E}_u$ and $\mathbf{E}_i$ through the existing user-item interaction matrix $\mathbf{R}$: 
\begin{equation}
    \text{RecSys}(\mathbf{R}, \mathbf{E}_u, \mathbf{E}_i) \rightarrow \hat{\mathbf{R}} \quad \forall u \in \mathcal{U}, i \in \mathcal{I}.
\end{equation}
In any multimodal recommendation setting (e.g., fashion, song, micro-video, or food recommendation), items' representation may be suitably enriched by considering the multimodal content describing them, such as products' images, descriptions, or audio tracks. Thus, let $\mathcal{M}$ be the set of modalities, and $\mathbf{F} \in \mathbb{R}^{|\mathcal{I}| \times |\mathcal{M}| \times c}$ be the multimodal feature tensor of all items, where $\mathbf{F}_{im} \in \mathbb{R}^{c}$ is the feature of item $i$ accounting for the $m$ modality (e.g., the visual feature extracted from the product image of item $i$ through a pre-trained deep convolutional network~\cite{DBLP:conf/cvpr/HeZRS16}). 

On such preliminaries, the multimodal recommendation task is about reconstructing the user-item interaction matrix $\hat{\mathbf{R}}$ by learning meaningful user and item embeddings $\mathbf{E}_u$ and $\mathbf{E}_i$ through the existing user-item interaction matrix $\mathbf{R}$ and the multimodal items' features $\mathbf{F}_i \in \mathbb{R}^{|\mathcal{M}| \times c}$:
\begin{equation}
\label{eq:mmrecsys}
    \text{MMRecSys}(\mathbf{R}, \mathbf{E}_u, \mathbf{E}_i, \mathbf{F}_i) \rightarrow \hat{\mathbf{R}} \quad \forall u \in \mathcal{U}, i \in \mathcal{I}.
\end{equation}

\begin{algorithm}[!t]
\caption{FeatProp for missing node features}
\label{alg:feat_prop}
\begin{algorithmic}[1]
\Input Feature matrix $\mathbf{X} \in \mathbb{R}^{|\mathcal{V}| \times k}$, feature mask $\mathbf{B} \in \mathbb{R}^{|\mathcal{V}| \times k}$, normalised adjacency matrix $\tilde{\mathbf{A}} \in \mathbb{R}^{|\mathcal{V}| \times |\mathcal{V}|}$.
\vspace{1mm}
\hrule
\vspace{1mm}
\State $\mathbf{Y} \gets \mathbf{X}$
\While{$\mathbf{X}$ has not converged}
\State $\mathbf{X} \gets \tilde{\mathbf{A}}\mathbf{X}$   \Comment{Propagate features}
\State $\mathbf{X}[\mathbf{B}] \gets \mathbf{Y}[\mathbf{B}]$                 \Comment{Reset known features}
\EndWhile
\vspace{1mm}
\hrule
\vspace{1mm}
\Output Imputed node features $\mathbf{X}$.
\end{algorithmic}
\end{algorithm}

\subsection{Missing node features in graph learning}
\label{sec:miss_node_feat_gnn}
Given an undirected graph $\mathcal{G}=(\mathcal{V}, \mathcal{E})$ with vertex and edge sets $\mathcal{V}$ and $\mathcal{E}$ respectively, its adjacency matrix $\mathbf{A} \in \mathbb{R}^{|\mathcal{V}| \times |\mathcal{V}|}$ is defined as $\mathbf{A}_{ij}=1$ if $(i,j) \in \mathcal{E}$, 0 otherwise. We let $\mathbf{D}=\mathrm{diag}(d_i)$ be the degree matrix and $\tilde{\mathbf{A}}=\mathbf{D}^{-\frac{1}{2}}\mathbf{A}\mathbf{D}^{-\frac{1}{2}}$ be the symmetrically normalised adjacency matrix. Graph neural networks (GNNs) models typically assume a fully observed feature matrix $\mathbf{X}\in \mathbb{R}^{|\mathcal{V}| \times k}$, where rows represent nodes and columns feature channels. 
Graph learning then aims at learning a function $f_\Theta(\mathcal{G}, \mathbf{X})$ to solve the task at hand, usually using graph neural networks~\cite{kipf2016gcn, velickovic2018graph, graphsage}.

However, in real-world scenarios, each feature is often only observed for a subset of the nodes. In that case, we additionally have a binary mask $\mathbf{B} \in \mathbb{R}^{|\mathcal{V}| \times k}$, such that $\mathbf{B}_{ij}$ is $1$ if feature $j$ of node $i$ is known, or $0$ otherwise. Several methods have been proposed for learning on graphs with missing node features, including SAT~\cite{sat}, GCNMF~\cite{taguchi2021gcnmf} PaGNN~\cite{jiang2021incomplete}, PCFI~\cite{um2023confidencebased}, and FeatProp~\cite{DBLP:conf/log/RossiK0C0B22}. 

FeatProp~\cite{DBLP:conf/log/RossiK0C0B22} (short for feature propagation) imputes the missing features by iteratively propagating the observed features in the graph, as shown in~\Cref{alg:feat_prop}. Running this iterative process to convergence minimizes the Dirichlet Energy of the graph~\cite{DBLP:conf/log/RossiK0C0B22}, a quantity measuring the smoothness of the graph features. As such, FeatProp is a fixed, non-learnable, pre-processing step after which a standard graph neural network can be run on the graph with the reconstructed features to solve the task at hand.

\section{Methodology}
In this section, we propose to tackle the problem of missing modalities in multimodal recommendation by reshaping it as a problem of missing node features in graph learning~\cite{DBLP:conf/log/RossiK0C0B22}. First, we introduce and formalise the missing modalities setting in multimodal recommendation. Then, we introduce our technique (inspired by the FeatProp algorithm~\cite{DBLP:conf/log/RossiK0C0B22}) and show how to integrate it as pre-processing phase in any multimodal recommender system. Finally, we summarise the differences between the original version of the FeatProp algorithm and our proposed one.

\subsection{Missing modalities in multimodal recommendation}
Differently from the missing node features issue as defined in graph learning (\Cref{sec:miss_node_feat_gnn}), the problem of missing modalities in multimodal recommendation implies the unavailability of the \textbf{whole multimodal feature vectors} for a specific subset of items from the catalogue. Realistically speaking, it is more likely some items are missing their multimodal content (e.g., product images are not available) to extract the multimodal features from instead of only some single vector entries of the extracted multimodal features. 

In this sense, we control the percentage of missing modalities, to be intended as the \textbf{percentage of items from the catalogue having no multimodal features}\footnote{In the remaining of the paper, when no confusion arises, we will refer to ``missing modalities'', ``missing multimodal features'', and ``missing items''  interchangeably.}. In the considered missing modalities setting, we indicate with $\mathbf{B} \in \mathbb{R}^{|\mathcal{I}| \times c}$ the binary mask matrix such that $\mathbf{B}_i = [1, 1, \dots, 1] \in \mathbb{R}^{c}$ if \textbf{all multimodal features} of item $i$ are available, $[0, 0, \dots, 0] \in \mathbb{R}^{c}$ otherwise. For instance, in a double modalities setting (e.g., visual and textual) we assume that any item $i$ with $\mathbf{B}_i = [0, 0, \dots, 0]$  does not have either visual or textual multimodal features. 

\subsection{FeatProp for multimodal recommendation}
\label{sec:feat_prop_for_mm_recsys}
After having outlined the setting of missing modalities in multimodal recommendation, we reshape the problem to suit the similar issue of missing node features in graph learning, eventually leveraging the FeatProp algorithm to address it~\cite{DBLP:conf/log/RossiK0C0B22}. 

To begin with, we observe that in the original work~\cite{DBLP:conf/log/RossiK0C0B22} the approach is applied on graphs connecting single-type nodes (i.e., monopartite graphs); conversely, our multimodal recommendation scenario implies a bipartite graph of users' and items' nodes. To translate the user-item bipartite graph to a monopartite setting, we consider the following assumption:

\begin{assumption}
    In a multimodal recommendation scenario, it is safe to assume that co-interacted items (i.e., items being interacted by the same users) may be characterised by similar multimodal content.
\end{assumption}

\noindent For instance, if the same users have clicked two fashion products (e.g., t-shirts), they likely have similar visual patterns (e.g., they are all in red shades) and are described with similar words. This assumption is also intuitively supported by the definition of collaborative filtering~\cite{DBLP:journals/fthci/EkstrandRK11} in recommender systems: similar users are likely to interact with similar items (and vice versa).  On an analogous side, the latest trends in multimodal recommendation propose to exploit the similarities between items' multimodal features as a form of pre-processing before the actual recommendation stage~\cite{DBLP:conf/mm/Zhang00WWW21,DBLP:conf/mm/LiuYLWTZSM21,DBLP:conf/mir/LiuMSO022, DBLP:conf/mm/ZhouS23}, which further confirm our assumption. 

\begin{figure*}[!t]
\centering

\subfloat[FeatProp as proposed in~\cite{DBLP:conf/log/RossiK0C0B22}]{
    \includegraphics[width=0.7\textwidth]{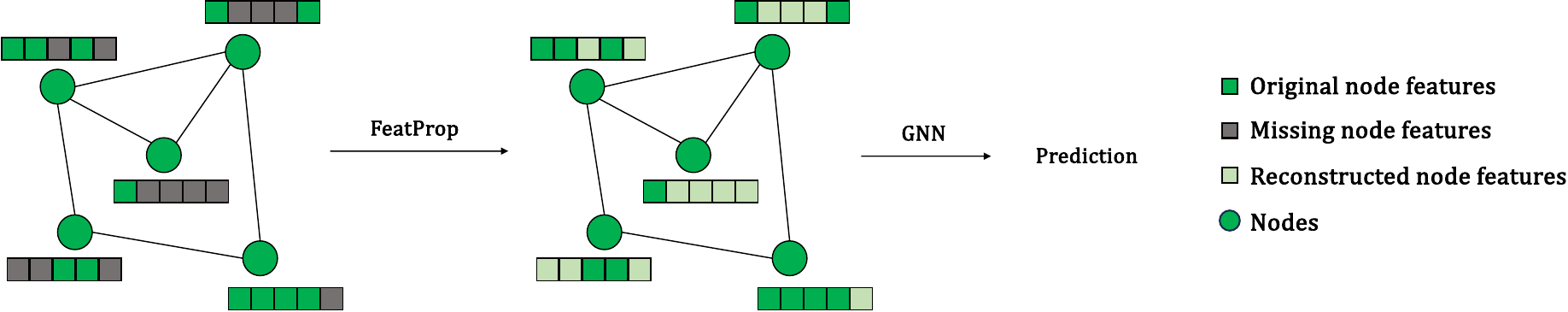}
}

\vspace{5mm}

\subfloat[Our FeatProp]{
    \includegraphics[width=0.7\textwidth]{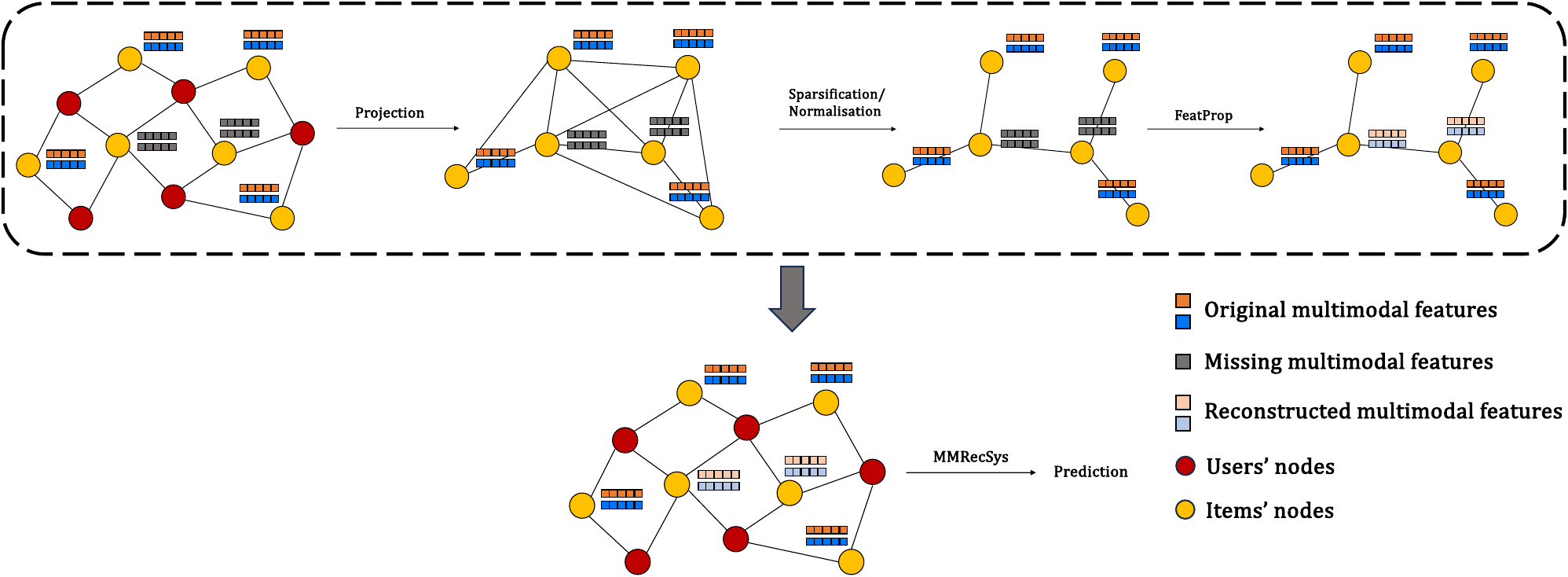}
}

\caption{The two versions of the FeatProp algorithm, as proposed (a) in~\cite{DBLP:conf/log/RossiK0C0B22} for missing node features, and (b) in this paper for missing modalities in multimodal recommendation. As for (b), the user-item graph is first projected into the item-item co-interaction graph; second, the graph is processed through sparsification and normalisation; then, the FeatProp algorithm is applied; finally, the reconstructed multimodal features are used as inputs to power any multimodal recommender system.}
\label{fig:method}
\end{figure*}

To obtain the graph representing the co-interactions among items, we first project the user-item graph on the item nodes, ending up with the \textbf{item-item projected graph}:
\begin{equation}
    \mathbf{R}^{\mathcal{I}} = \mathbf{R}^\top \mathbf{R},
\end{equation}
where $\mathbf{R}^{\mathcal{I}} \in \mathbb{R}^{|\mathcal{I}| \times |\mathcal{I}|}$ is the item-item co-interaction matrix, and $\mathbf{R}^{\mathcal{I}}_{ij}$ is the number of users who interacted with both items $i$ and $j$. As the projected item-item graph may be too dense (i.e., with noisy item-item interactions) we perform the sparsification of the item-item matrix through top-$n$ sparsification (compare with~\cite{DBLP:conf/mm/Zhang00WWW21}):
\begin{equation}
    \overline{\mathbf{R}}_{ij}^{\mathcal{I}} = \text{sparse}\left(\mathbf{R}^{\mathcal{I}}, n\right) = 
    \begin{cases}
        1 & \text{if } \mathbf{R}_{ij}^{\mathcal{I}} \in \text{top-}n\left(\mathbf{R}^{\mathcal{I}}_i\right) \\
        0 & \text{otherwise},
    \end{cases}
\end{equation}
where $\overline{\mathbf{R}}^{\mathcal{I}}$ is the sparsified item-item matrix, while $\text{top-}n(\cdot)$ is the function returning the $n$ highest values of a matrix row-wise, with $n$ sparsification rate. Furthermore, we also normalise the item-item co-interaction matrix through the symmetric Laplacian normalisation:
\begin{equation} \tilde{\mathbf{R}}^{\mathcal{I}} = \left(\mathbf{D}^{\mathcal{I}}\right)^{-\frac{1}{2}}\mathbf{\overline{R}}^{\mathcal{I}}\left(\mathbf{D}^{\mathcal{I}}\right)^{-\frac{1}{2}},
\end{equation}
where $\mathbf{D}^{\mathcal{I}} \in \mathbb{R}^{|\mathcal{I}| \times |\mathcal{I}|}$ is the degree matrix of the item-item co-interactions matrix. Then, we can apply the FeatProp algorithm introduced above (\Cref{alg:feat_prop}) to impute the missing multimodal features; note that the algorithm will be iterated over the set of modalities $\mathcal{M}$ as in our setting \textbf{each missing multimodal feature can be only imputed by the available multimodal features of the same modality} 
Finally, we inject the reconstructed multimodal features along with the original (existing) ones into any multimodal recommender system to train it and approximate the user-item interaction matrix (as already defined in~\Cref{eq:mmrecsys}).
Our FeatProp algorithm is reported in~\Cref{alg:feat_prop_ours}, where $\mathbf{F}_{\mathcal{I}m} \in \mathbb{R}^{|\mathcal{I}| \times c}$ are the features for all items considering the single modality $m$. Moreover, we report a schematic representation of the method, along with the original FeatProp, in~\Cref{fig:method}.

\begin{algorithm}[!t]
\caption{FeatProp for missing modalities}
\label{alg:feat_prop_ours}
\begin{algorithmic}[1]
\Input Multimodal features of all items $\mathbf{F}_{\mathcal{I}m} \in \mathbb{R}^{|\mathcal{I}| \times c}$, feature mask $\mathbf{B} \in \mathbb{R}^{|\mathcal{I}| \times c}$, item-item co-interaction matrix $\mathbf{R}^{\mathcal{I}} \in \mathbb{R}^{|\mathcal{I}| \times|\mathcal{I}|}$, sparsification rate $n$, set of modalities $\mathcal{M}$.\\
\vspace{1mm}
\hrule
\vspace{1mm}
$\mathbf{R}^{\mathcal{I}} = \mathbf{R}^\top \mathbf{R}$ \Comment{Project on the item-item graph} \\
$\overline{\mathbf{R}}^{\mathcal{I}} = \text{sparse}\left(\mathbf{R}^{\mathcal{I}}, n\right)$ \Comment{Perform top-$n$ sparsification} \\
$\tilde{\mathbf{R}}^{\mathcal{I}} = \left(\mathbf{D}^{\mathcal{I}}\right)^{-\frac{1}{2}}\mathbf{\overline{R}}^{\mathcal{I}}\left(\mathbf{D}^{\mathcal{I}}\right)^{-\frac{1}{2}}$ \Comment{Adjacency normalisation}
\For{$m \in \mathcal{M}$} \Comment{Repeat for all modalities}
\State $\mathbf{Y} \gets \mathbf{F}_{\mathcal{I}m}$
\While{$\mathbf{F}_{\mathcal{I}m}$ has not converged}
\State $\mathbf{F}_{\mathcal{I}m} \gets \tilde{\mathbf{R}}^{\mathcal{I}}\mathbf{F}_{\mathcal{I}m}$   \Comment{Propagate features}
\State $\mathbf{F}_{\mathcal{I}m}[\mathbf{B}] \gets \mathbf{Y}[\mathbf{B}]$                 \Comment{Reset known features}
\EndWhile
\EndFor
\vspace{1mm}
\hrule
\vspace{1mm}
\Output Imputed missing multimodal features $\mathbf{F}$.
\end{algorithmic}
\end{algorithm}

\subsection{Differences between the original and proposed FeatProp}
This last part summarises the important differences underlying the originally-proposed FeatProp algorithm~\cite{DBLP:conf/log/RossiK0C0B22} and our version, categorising them into \textbf{task}, \textbf{graph topology}, and \textbf{node features}.

\begin{enumerate}[leftmargin=*]
\item \textbf{Task.} In the original work~\cite{DBLP:conf/log/RossiK0C0B22}, the authors take into account classical machine learning tasks on graphs such as node classification. Conversely, in this work, we are performing personalised recommendation exploiting the multimodal content of items.

\item \textbf{Graph topology.} The authors in~\cite{DBLP:conf/log/RossiK0C0B22} work on graphs having only one type of nodes (i.e., monopartite graphs). Differently from such a setting, our approach works with user-item graphs, which are (by design) bipartite.

\item \textbf{Node features.} The assumption made in~\cite{DBLP:conf/log/RossiK0C0B22} is that, in real-world scenarios, node features come in the form of attribute vectors, and they may be missing some entries (i.e., attributes). In our setting, instead, this assumption would not be realistic, since multimodal features come from pre-trained deep learning models, so they are (by-nature) not interpretable embeddings. Thus, it would not make sense to consider a setting where only some multimodal features entries are missing; conversely, it is more realistic if the whole feature vector is not available.
\end{enumerate}
\section{Experimental settings}
This section provides details regarding the experimental settings for this work. First, we present the adopted recommendation datasets. Second, we briefly introduce the selected multimodal recommender systems we use in conjunction with our proposed FeatProp technique. Then, we describe the step to simulate the missing modalities setting, along with the other baselines we compare our approach against. Finally, we outline the evaluation metrics and useful information to fully reproduce our experiments. 

\subsection{Datasets}
We select three popular recommendation datasets generally adopted in multimodal recommendation. Specifically, we consider three sub-categories from the Amazon catalogue~\cite{DBLP:conf/sigir/McAuleyTSH15}, namely, \textbf{Amazon Baby} and \textbf{Amazon Sports} (in the exact versions reported in~\cite{DBLP:conf/www/WeiHXZ23}) along with \textbf{Amazon Toys} (considering the version used in~\cite{DBLP:journals/corr/abs-2309-05273}). The items in all three datasets come with both visual and textual features and are 4096- and 1024-dimensional vectors, where the former are publicly available at this URL\footnote{\url{https://cseweb.ucsd.edu/~jmcauley/datasets/amazon/links.html}.}, while the latter are obtained by concatenating item's product title, description, brand, and categorical information and using SentenceBert~\cite{DBLP:conf/emnlp/ReimersG19} for the multimodal extraction. Datasets' statistics are reported in~\Cref{tab:datasets}.

\subsection{Multimodal recommender systems}
To test the effectiveness of our FeatProp approach, we use it as pre-processing module on top of two recent multimodal recommender systems: \texttt{MMSSL}~\cite{DBLP:conf/www/WeiHXZ23} and \texttt{FREEDOM}~\cite{DBLP:conf/mm/ZhouS23}. Multimodal self-supervised learning (\texttt{MMSSL})~\cite{DBLP:conf/www/WeiHXZ23} for recommendation performs an interactive structure learning task which uses adversarial perturbations to seek data augmentation, thus characterising the interdependencies between the user-item graph and the items' multimodal semantics. Furthermore, the authors propose to address an additional cross-modality contrastive learning task to jointly learn inter-modalities similarities and preserve the diversity of the users' preferences towards such modalities. Inspired by the state-of-the-art multimodal recommendation approach \texttt{LATTICE}~\cite{DBLP:conf/mm/Zhang00WWW21}, the authors from~\cite{DBLP:conf/mm/ZhouS23} observe how freezing the item-item multimodal structure as a form of pre-training can lead to improved recommendation performance. Additionally, they also propose to jointly denoise the user-item interaction graph by pruning noisy interactions; they name their method \texttt{FREEDOM}, short for method ``that freezes the item-item graph and denoises the user-item interaction graph simultaneously for multimodal recommendation''.

\begin{table}[!t]
    \caption{Statistics of the tested datasets.}\label{tab:datasets}
    \centering
    \begin{tabular}{lrrrc}
    \toprule
        \textbf{Datasets} & \textbf{\# Users} & \textbf{\# Items} & \textbf{\# Interactions} & \textbf{Sparsity (\%)}\\ \cmidrule{1-5}
        Amazon Baby & 19,445 & 7,050 & 139,110 & 99.899 \\
        Amazon Toys & 19,412 & 11,924 & 167,597 & 99.928 \\ 
        Amazon Sports & 35,598 & 18,357 & 256,308 & 99.961 \\
        \bottomrule
    \end{tabular}
\end{table}

\subsection{Missing modalities settings}
\label{sec:miss_modes}
We consider the settings where the $[10\%, 20\%, \dots, 90\%]$ of items in the catalogue have no multimodal features. To simulate each scenario, we randomly select the missing items from the initial set with uniform distribution; the task is to impute the multimodal features of the missing items through a pre-training solution in such a way the final recommendation performance is degraded the least. For each setting of percentage, recommendation model, and dataset, \textbf{we repeat the experiments 5 times with 5 differently sampled item sets} to ensure that the obtained results are not dependent on the specific items that have been randomly selected. We compare the \textbf{FeatProp} algorithm with three shallow techniques usually adopted for missing values in machine learning, namely: (i) \textbf{Zeros}, where the multimodal features of missing items are replaced with zeros-valued vectors, (ii) \textbf{Mean}, where the multimodal features of missing items are replaced with the global mean of the multimodal features of the non-missing items, and (iii) \textbf{Random}, where the multimodal features of the missing items are replaced with random-valued vectors. Overall, \textbf{we present an extensive experimental setting of 1080 different runnings} comprising 4 methods to impute missing modalities, 3 recommendation datasets, 2 multimodal recommender systems, 9 percentages of missing items, and 5 random samplings of missing items.  

\subsection{Reproducibility}
Datasets are split into train, validation, and test sets as in~\cite{DBLP:conf/www/WeiHXZ23}. We calculate the Recall@20 on the validation set to perform the early stopping strategy to greatly reduce the computational time required to run all the 1080 settings described above; the same metric is used to assess the models' performance on the test set. Note that we purposely decide not to perform the hyper-parameter exploration because our goal is not to compare the recommender systems, but rather to calculate the performance variation when using, for each recommendation approach, different strategies for imputing missing modalities \textbf{at the same starting hyper-parameter settings}. Since for \texttt{MMSSL} our experimental settings are the same as in the original work, we decide to follow the best hyper-parameter values\footnote{\url{https://github.com/HKUDS/MMSSL/issues/10}.}; conversely, as the authors of \texttt{FREEDOM} use different splittings and/or datasets with respect to the ones we use, we follow the best hyper-parameter values reported in a recent survey on multimodal recommendation~\cite{DBLP:journals/corr/abs-2309-05273}. Regarding FeatProp, we fix the sparsification rate $n = 20$ as we empirically find this provides high performance, and set the convergence criterion for the features propagation through a maximum number of propagation layers $L \in [1, 2, 3, 20]$. While with few propagation layers we try to understand whether FeatProp can rapidly converge to sufficient recommendation performance, we also investigate the $L = 20$ setting to align with the original FeatProp work. 
\section{Results and discussion}
In this section, we present the results regarding the performance of FeatProp to mitigate the missing modalities issue in multimodal recommendation. Specifically, we first describe the overall performance of the approach when compared to other shallower solutions to address missing values in machine learning. Then, to assess the effectiveness of the architectural choices behind FeatProp, we conduct a parameter study to evaluate the impact of propagation layers.

\subsection{Overall performance}
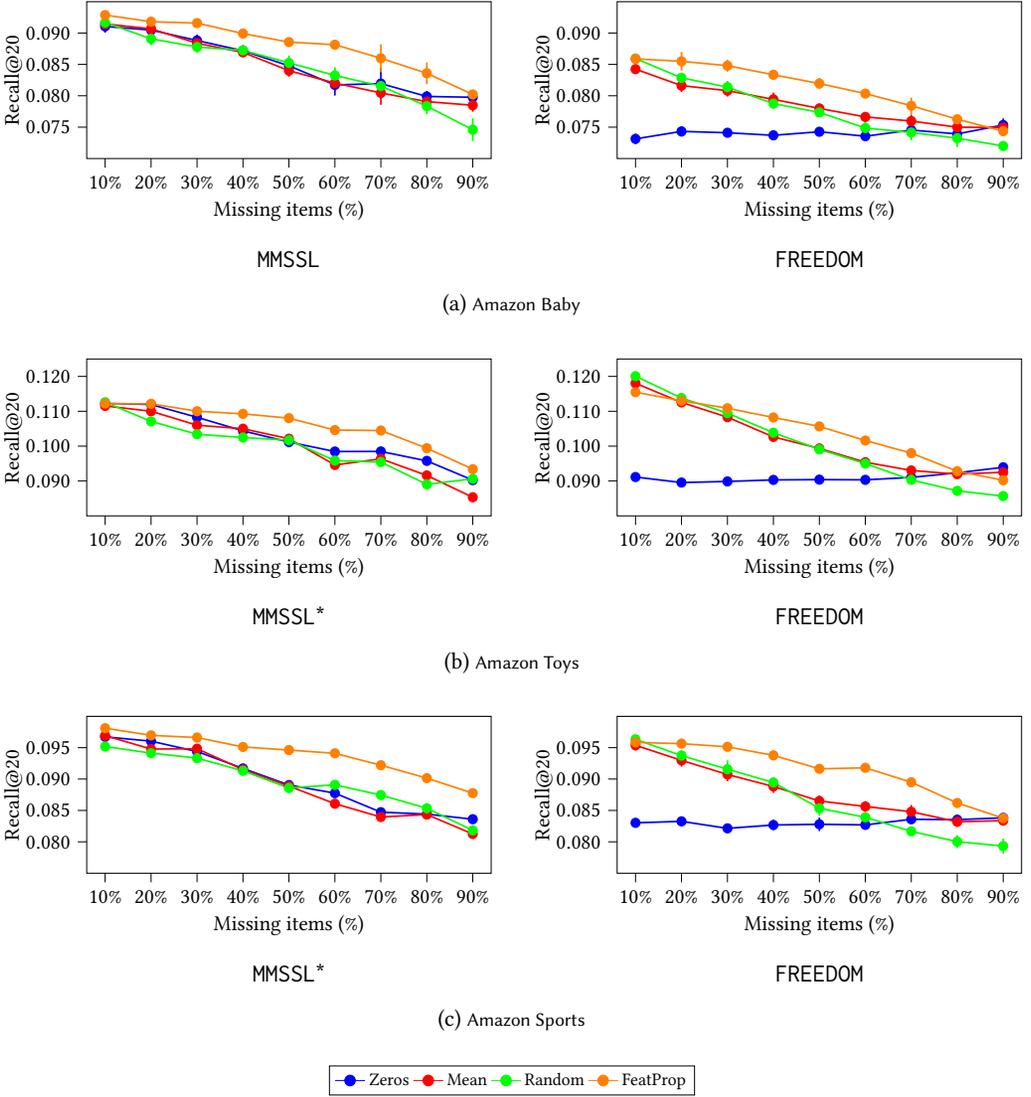
\begin{figure*}[!t]
\captionsetup[subfigure]{font=large,labelfont=large}
\centering

\begin{adjustbox}
{width=\textwidth,center}
\subfloat[Amazon Baby]{
    \begin{tikzpicture}

\begin{axis}[
    yticklabel style={
    /pgf/number format/.cd,
    fixed, fixed zerofill,
    precision=3
    },
    width=0.6\textwidth,
    height=0.3\textwidth,
    scaled y ticks=false,
    ylabel=Recall@20,
    xlabel=\makecell{Missing items (\%)\\\phantom{ }\\\Large \texttt{MMSSL}},
    tick align=outside,
    tick pos=left,
    x grid style={darkgray176},
    xmin=6, xmax=94,
    ymin=0.070, ymax=0.095,
    xtick style={color=black},
    xticklabels={10\%, 20\%, 30\%, 40\%, 50\%, 60\%, 70\%, 80\%, 90\%},
    xtick={10, 20, 30, 40, 50, 60, 70, 80, 90},
    ytick={0.075, 0.080, 0.085, 0.090},
    y grid style={darkgray176},
    ytick style={color=black}
]


\path [draw=blue, thick]
(axis cs:10,0.0900482981903004)
--(axis cs:10,0.0920554618096996);

\path [draw=blue, thick]
(axis cs:20,0.0900624025322211)
--(axis cs:20,0.0908914934677789);

\path [draw=blue, thick]
(axis cs:30,0.0878659058407647)
--(axis cs:30,0.0897516941592353);

\path [draw=blue, thick]
(axis cs:40,0.086282044945072)
--(axis cs:40,0.088073755054928);

\path [draw=blue, thick]
(axis cs:50,0.084114067727236)
--(axis cs:50,0.085438788272764);

\path [draw=blue, thick]
(axis cs:60,0.0800378113849816)
--(axis cs:60,0.0834164686150184);

\path [draw=blue, thick]
(axis cs:70,0.0794746806952489)
--(axis cs:70,0.0844150233047511);

\path [draw=blue, thick]
(axis cs:80,0.0790065971871822)
--(axis cs:80,0.0807754028128178);

\path [draw=blue, thick]
(axis cs:90,0.0790385808221942)
--(axis cs:90,0.0804822631778058);

\addplot [thick, blue, forget plot, mark=*]
table {%
10 0.09105188
20 0.090476948
30 0.0888088
40 0.0871779
50 0.084776428
60 0.08172714
70 0.081944852
80 0.079891
90 0.079760422
};


\path [draw=red, thick]
(axis cs:10,0.0904798426380837)
--(axis cs:10,0.0924131493619163);

\path [draw=red, thick]
(axis cs:20,0.09023076086705)
--(axis cs:20,0.09111060313295);

\path [draw=red, thick]
(axis cs:30,0.0881572784468995)
--(axis cs:30,0.0885952735531005);

\path [draw=red, thick]
(axis cs:40,0.0861946271532065)
--(axis cs:40,0.0876328488467934);

\path [draw=red, thick]
(axis cs:50,0.0830381025108237)
--(axis cs:50,0.0849661094891763);

\path [draw=red, thick]
(axis cs:60,0.0807992044583998)
--(axis cs:60,0.0832544755416002);

\path [draw=red, thick]
(axis cs:70,0.0785609343456745)
--(axis cs:70,0.0823490296543255);

\path [draw=red, thick]
(axis cs:80,0.0775687547390399)
--(axis cs:80,0.0805662052609601);

\path [draw=red, thick]
(axis cs:90,0.0774519332402091)
--(axis cs:90,0.0795310587597909);

\addplot [thick, red, forget plot, mark=*]
table {%
10 0.091446496
20 0.090670682
30 0.088376276
40 0.086913738
50 0.084002106
60 0.08202684
70 0.080454982
80 0.07906748
90 0.078491496
};

\path [draw=green, thick]
(axis cs:10,0.090629738558373)
--(axis cs:10,0.0926891214416271);

\path [draw=green, thick]
(axis cs:20,0.0880481901764048)
--(axis cs:20,0.0901165458235951);

\path [draw=green, thick]
(axis cs:30,0.0868856090163769)
--(axis cs:30,0.0886628069836231);

\path [draw=green, thick]
(axis cs:40,0.0865533377255216)
--(axis cs:40,0.0879337742744784);

\path [draw=green, thick]
(axis cs:50,0.0840939372782643)
--(axis cs:50,0.0864163947217357);

\path [draw=green, thick]
(axis cs:60,0.0819276723555408)
--(axis cs:60,0.0845479396444592);

\path [draw=green, thick]
(axis cs:70,0.080871834117455)
--(axis cs:70,0.082235045882545);

\path [draw=green, thick]
(axis cs:80,0.0770686722789399)
--(axis cs:80,0.0796192717210601);

\path [draw=green, thick]
(axis cs:90,0.0728036552475859)
--(axis cs:90,0.0763786287524141);

\addplot [thick, green, forget plot, mark=*]
table {%
10 0.09165943
20 0.089082368
30 0.087774208
40 0.087243556
50 0.085255166
60 0.083237806
70 0.08155344
80 0.078343972
90 0.074591142
};


\path [draw=orange, thick]
(axis cs:10,0.0923344654564221)
--(axis cs:10,0.0933875625435779);

\path [draw=orange, thick]
(axis cs:20,0.0912315090740903)
--(axis cs:20,0.0923787709259097);

\path [draw=orange, thick]
(axis cs:30,0.0908831387259672)
--(axis cs:30,0.0922787932740328);

\path [draw=orange, thick]
(axis cs:40,0.0891369732761806)
--(axis cs:40,0.0906941027238194);

\path [draw=orange, thick]
(axis cs:50,0.0878951038198811)
--(axis cs:50,0.0892147881801189);

\path [draw=orange, thick]
(axis cs:60,0.0874789659651524)
--(axis cs:60,0.0887955580348475);

\path [draw=orange, thick]
(axis cs:70,0.0837641212647471)
--(axis cs:70,0.0881989987352529);

\path [draw=orange, thick]
(axis cs:80,0.0818528595218961)
--(axis cs:80,0.0853128444781039);

\path [draw=orange, thick]
(axis cs:90,0.0795941096303174)
--(axis cs:90,0.0808305543696826);

\addplot [thick, orange, forget plot, mark=*]
table {%
10 0.092861014
20 0.09180514
30 0.091580966
40 0.089915538
50 0.088554946
60 0.088137262
70 0.08598156
80 0.083582852
90 0.080212332
};

\end{axis}

\end{tikzpicture} \quad
    \begin{tikzpicture}

\begin{axis}[
    yticklabel style={
    /pgf/number format/.cd,
    fixed, fixed zerofill,
    precision=3
    },
    width=0.6\textwidth,
    height=0.3\textwidth,
    scaled y ticks=false,
    ylabel=Recall@20,
    xlabel=\makecell{Missing items (\%)\\\phantom{ }\\\Large \texttt{FREEDOM}},
    tick align=outside,
    tick pos=left,
    x grid style={darkgray176},
    xmin=6, xmax=94,
    ymin=0.070, ymax=0.095,
    xtick style={color=black},
    xticklabels={10\%, 20\%, 30\%, 40\%, 50\%, 60\%, 70\%, 80\%, 90\%},
    xtick={10, 20, 30, 40, 50, 60, 70, 80, 90},
    ytick={0.075, 0.080, 0.085, 0.090},
    y grid style={darkgray176},
    ytick style={color=black}
]


\path [draw=blue, thick]
(axis cs:10,0.0723119101178591)
--(axis cs:10,0.0739536612274365);

\path [draw=blue, thick]
(axis cs:20,0.073509081210457)
--(axis cs:20,0.0751362178720399);

\path [draw=blue, thick]
(axis cs:30,0.0735052223094519)
--(axis cs:30,0.0747131659614001);

\path [draw=blue, thick]
(axis cs:40,0.073139162667256)
--(axis cs:40,0.0742579795040863);

\path [draw=blue, thick]
(axis cs:50,0.0738544900160982)
--(axis cs:50,0.0746793664850456);

\path [draw=blue, thick]
(axis cs:60,0.0729758461568241)
--(axis cs:60,0.0741357090917541);

\path [draw=blue, thick]
(axis cs:70,0.0740599444298829)
--(axis cs:70,0.0750066321030036);

\path [draw=blue, thick]
(axis cs:80,0.0734587691016261)
--(axis cs:80,0.0743870825401313);

\path [draw=blue, thick]
(axis cs:90,0.0742877681796947)
--(axis cs:90,0.0763938685321859);

\addplot [thick, blue, forget plot, mark=*]
table {%
10 0.0731327856726478
20 0.0743226495412485
30 0.074109194135426
40 0.0736985710856712
50 0.0742669282505719
60 0.0735557776242891
70 0.0745332882664433
80 0.0739229258208787
90 0.0753408183559403
};


\path [draw=red, thick]
(axis cs:10,0.0836335433699559)
--(axis cs:10,0.0848525765177687);

\path [draw=red, thick]
(axis cs:20,0.0805918674236017)
--(axis cs:20,0.0826466294976439);

\path [draw=red, thick]
(axis cs:30,0.0798829815701438)
--(axis cs:30,0.0817455007043626);

\path [draw=red, thick]
(axis cs:40,0.0783387217589581)
--(axis cs:40,0.0805185386639848);

\path [draw=red, thick]
(axis cs:50,0.0772032786927799)
--(axis cs:50,0.0787617160902377);

\path [draw=red, thick]
(axis cs:60,0.076043914236175)
--(axis cs:60,0.0772175570018419);

\path [draw=red, thick]
(axis cs:70,0.0749355092991127)
--(axis cs:70,0.0770087351201858);

\path [draw=red, thick]
(axis cs:80,0.0743765638343385)
--(axis cs:80,0.0756113856711504);

\path [draw=red, thick]
(axis cs:90,0.0741653854331191)
--(axis cs:90,0.0757187632680237);

\addplot [thick, red, forget plot, mark=*]
table {%
10 0.0842430599438623
20 0.0816192484606228
30 0.0808142411372532
40 0.0794286302114715
50 0.0779824973915088
60 0.0766307356190084
70 0.0759721222096492
80 0.0749939747527444
90 0.0749420743505714
};


\path [draw=green, thick]
(axis cs:10,0.0853476334851278)
--(axis cs:10,0.0865186501129735);

\path [draw=green, thick]
(axis cs:20,0.0823820555715047)
--(axis cs:20,0.0833352917521589);

\path [draw=green, thick]
(axis cs:30,0.0804207849441061)
--(axis cs:30,0.0823529596735846);

\path [draw=green, thick]
(axis cs:40,0.0779795290512049)
--(axis cs:40,0.0795320928179933);

\path [draw=green, thick]
(axis cs:50,0.0768562435849089)
--(axis cs:50,0.0778524136666966);

\path [draw=green, thick]
(axis cs:60,0.0736376404609324)
--(axis cs:60,0.0760784136389299);

\path [draw=green, thick]
(axis cs:70,0.0729402850808958)
--(axis cs:70,0.0753617708510336);

\path [draw=green, thick]
(axis cs:80,0.071852179967902)
--(axis cs:80,0.0746307411503448);

\path [draw=green, thick]
(axis cs:90,0.0717617906507796)
--(axis cs:90,0.0722363282964186);

\addplot [thick, green, forget plot, mark=*]
table {%
10 0.0859331417990506
20 0.0828586736618318
30 0.0813868723088454
40 0.0787558109345991
50 0.0773543286258028
60 0.0748580270499312
70 0.0741510279659647
80 0.0732414605591234
90 0.0719990594735991
};


\path [draw=orange, thick]
(axis cs:10,0.0850815231000443)
--(axis cs:10,0.0866768693835822);

\path [draw=orange, thick]
(axis cs:20,0.0840103458651102)
--(axis cs:20,0.0869828402170585);

\path [draw=orange, thick]
(axis cs:30,0.0838428513308321)
--(axis cs:30,0.0857543143536703);

\path [draw=orange, thick]
(axis cs:40,0.082636637207996)
--(axis cs:40,0.0840572900285709);

\path [draw=orange, thick]
(axis cs:50,0.0810559566091546)
--(axis cs:50,0.0828524694297107);

\path [draw=orange, thick]
(axis cs:60,0.0799623437108186)
--(axis cs:60,0.0807322497319489);

\path [draw=orange, thick]
(axis cs:70,0.0770934125257779)
--(axis cs:70,0.0797286859257826);

\path [draw=orange, thick]
(axis cs:80,0.0756223640467286)
--(axis cs:80,0.0769013328098558);

\path [draw=orange, thick]
(axis cs:90,0.0736251525277622)
--(axis cs:90,0.0750832891865431);

\addplot [thick, orange, forget plot, mark=*]
table {%
10 0.0858791962418133
20 0.0854965930410844
30 0.0847985828422512
40 0.0833469636182834
50 0.0819542130194326
60 0.0803472967213837
70 0.0784110492257802
80 0.0762618484282922
90 0.0743542208571527
};

\end{axis}

\end{tikzpicture}
}
\end{adjustbox}

\vspace{5mm}

\begin{adjustbox}
{width=\textwidth,center}
\subfloat[Amazon Toys]{
    \begin{tikzpicture}

\begin{axis}[
    yticklabel style={
    /pgf/number format/.cd,
    fixed, fixed zerofill,
    precision=3
    },
    width=0.6\textwidth,
    height=0.3\textwidth,
    scaled y ticks=false,
    ylabel=Recall@20,
    xlabel=\makecell{Missing items (\%)\\\phantom{ }\\\Large \texttt{MMSSL}*},
    tick align=outside,
    tick pos=left,
    x grid style={darkgray176},
    xmin=6, xmax=94,
    xtick style={color=black},
    ymin=0.080, ymax=0.125,
    xticklabels={10\%, 20\%, 30\%, 40\%, 50\%, 60\%, 70\%, 80\%, 90\%},
    xtick={10, 20, 30, 40, 50, 60, 70, 80, 90},
    ytick={0.090, 0.100, 0.110, 0.120},
    y grid style={darkgray176},
    ytick style={color=black}
]


\addplot [thick, blue, forget plot, mark=*]
table {%
10 0.11228135
20 0.11194175
30 0.1082765
40 0.10437009
50 0.10117028
60 0.09845673
70 0.09847867
80 0.09577209
90 0.0902482
};


\addplot [thick, red, forget plot, mark=*]
table {%
10 0.11154067
20 0.1100125
30 0.10599603
40 0.10499418
50 0.10211568
60 0.09459787
70 0.09635341
80 0.09160058
90 0.08533306
};


\addplot [thick, green, forget plot, mark=*]
table {%
10 0.1125979
20 0.10710717
30 0.10341022
40 0.10247101
50 0.10169129
60 0.09585316
70 0.09542893
80 0.08903568
90 0.09061964
};


\addplot [thick, orange, forget plot, mark=*]
table {%
10 0.11219028
20 0.11214571
30 0.11001803
40 0.10926025
50 0.10802647
60 0.10461161
70 0.10447713
80 0.0994101
90 0.09339517
};

\end{axis}

\end{tikzpicture} \quad
    \begin{tikzpicture}

\begin{axis}[
    yticklabel style={
    /pgf/number format/.cd,
    fixed, fixed zerofill,
    precision=3
    },
    width=0.6\textwidth,
    height=0.3\textwidth,
    scaled y ticks=false,
    ylabel=Recall@20,
    xlabel=\makecell{Missing items (\%)\\\phantom{ }\\\Large \texttt{FREEDOM}},
    tick align=outside,
    tick pos=left,
    x grid style={darkgray176},
    xmin=6, xmax=94,
    ymin=0.080, ymax=0.125,
    xtick style={color=black},
    xticklabels={10\%, 20\%, 30\%, 40\%, 50\%, 60\%, 70\%, 80\%, 90\%},
    xtick={10, 20, 30, 40, 50, 60, 70, 80, 90},
    ytick={0.090, 0.100, 0.110, 0.120},
    y grid style={darkgray176},
    ytick style={color=black}
]


\path [draw=blue, thick]
(axis cs:10,0.0905574207534426)
--(axis cs:10,0.0917123509846939);

\path [draw=blue, thick]
(axis cs:20,0.0889795959323449)
--(axis cs:20,0.0901251728473088);

\path [draw=blue, thick]
(axis cs:30,0.0896376194884493)
--(axis cs:30,0.0901160730172704);

\path [draw=blue, thick]
(axis cs:40,0.0898992298839737)
--(axis cs:40,0.090724376638847);

\path [draw=blue, thick]
(axis cs:50,0.089362980805349)
--(axis cs:50,0.0914237959829583);

\path [draw=blue, thick]
(axis cs:60,0.0893642271497637)
--(axis cs:60,0.0913032506226415);

\path [draw=blue, thick]
(axis cs:70,0.0891521749243157)
--(axis cs:70,0.0929168107159995);

\path [draw=blue, thick]
(axis cs:80,0.0917438901152723)
--(axis cs:80,0.092940920052231);

\path [draw=blue, thick]
(axis cs:90,0.0933507732875926)
--(axis cs:90,0.0945040659949088);

\addplot [thick, blue, forget plot, mark=*]
table {%
10 0.0911348858690682
20 0.0895523843898269
30 0.0898768462528599
40 0.0903118032614103
50 0.0903933883941537
60 0.0903337388862026
70 0.0910344928201576
80 0.0923424050837516
90 0.0939274196412507
};


\path [draw=red, thick]
(axis cs:10,0.11702789286747)
--(axis cs:10,0.119037835528832);

\path [draw=red, thick]
(axis cs:20,0.111462871443931)
--(axis cs:20,0.113616551404717);

\path [draw=red, thick]
(axis cs:30,0.107494422203575)
--(axis cs:30,0.109112982636202);

\path [draw=red, thick]
(axis cs:40,0.101367995555859)
--(axis cs:40,0.103913025154945);

\path [draw=red, thick]
(axis cs:50,0.0987252386764935)
--(axis cs:50,0.0999606605990349);

\path [draw=red, thick]
(axis cs:60,0.0949214980402902)
--(axis cs:60,0.0958375146498094);

\path [draw=red, thick]
(axis cs:70,0.0918793499365413)
--(axis cs:70,0.0941722523959335);

\path [draw=red, thick]
(axis cs:80,0.0910897288379098)
--(axis cs:80,0.0928235910105812);

\path [draw=red, thick]
(axis cs:90,0.0916466930265704)
--(axis cs:90,0.0933778335571121);

\addplot [thick, red, forget plot, mark=*]
table {%
10 0.118032864198151
20 0.112539711424324
30 0.108303702419889
40 0.102640510355402
50 0.0993429496377642
60 0.0953795063450498
70 0.0930258011662374
80 0.0919566599242455
90 0.0925122632918412
};


\path [draw=green, thick]
(axis cs:10,0.119261586504341)
--(axis cs:10,0.120873662379928);

\path [draw=green, thick]
(axis cs:20,0.112457259497929)
--(axis cs:20,0.115138716300985);

\path [draw=green, thick]
(axis cs:30,0.108310050671528)
--(axis cs:30,0.110650065251054);

\path [draw=green, thick]
(axis cs:40,0.102888339253653)
--(axis cs:40,0.104797613807435);

\path [draw=green, thick]
(axis cs:50,0.0979824866876343)
--(axis cs:50,0.100149291518017);

\path [draw=green, thick]
(axis cs:60,0.0937453627799352)
--(axis cs:60,0.096262534446864);

\path [draw=green, thick]
(axis cs:70,0.0893831580437376)
--(axis cs:70,0.0912718803440659);

\path [draw=green, thick]
(axis cs:80,0.0863206636002525)
--(axis cs:80,0.0880130477584591);

\path [draw=green, thick]
(axis cs:90,0.0848228215300546)
--(axis cs:90,0.0865222899204741);

\addplot [thick, green, forget plot, mark=*]
table {%
10 0.120067624442135
20 0.113797987899457
30 0.109480057961291
40 0.103842976530544
50 0.0990658891028259
60 0.0950039486133996
70 0.0903275191939017
80 0.0871668556793558
90 0.0856725557252643
};


\path [draw=orange, thick]
(axis cs:10,0.114742816406687)
--(axis cs:10,0.116205920665581);

\path [draw=orange, thick]
(axis cs:20,0.112568757364267)
--(axis cs:20,0.113434020969854);

\path [draw=orange, thick]
(axis cs:30,0.110104201089019)
--(axis cs:30,0.111642976080683);

\path [draw=orange, thick]
(axis cs:40,0.10731342141478)
--(axis cs:40,0.109116383016775);

\path [draw=orange, thick]
(axis cs:50,0.104429778797549)
--(axis cs:50,0.106894833647252);

\path [draw=orange, thick]
(axis cs:60,0.100296726238646)
--(axis cs:60,0.102906275567408);

\path [draw=orange, thick]
(axis cs:70,0.0967488912444312)
--(axis cs:70,0.0992492043956919);

\path [draw=orange, thick]
(axis cs:80,0.092533419600392)
--(axis cs:80,0.0930672106501548);

\path [draw=orange, thick]
(axis cs:90,0.0899827868145382)
--(axis cs:90,0.0904578913742403);

\addplot [thick, orange, forget plot, mark=*]
table {%
10 0.115474368536134
20 0.113001389167061
30 0.110873588584851
40 0.108214902215777
50 0.105662306222401
60 0.101601500903027
70 0.0979990478200616
80 0.0928003151252734
90 0.0902203390943892
};

\end{axis}

\end{tikzpicture}
}
\end{adjustbox}

\vspace{5mm}

\begin{adjustbox}
{width=\textwidth,center}
\subfloat[Amazon Sports]{
    \begin{tikzpicture}

\begin{axis}[
    yticklabel style={
    /pgf/number format/.cd,
    fixed, fixed zerofill,
    precision=3
    },
    width=0.6\textwidth,
    height=0.3\textwidth,
    scaled y ticks=false,
    ylabel=Recall@20,
    xlabel=\makecell{Missing items (\%)\\\phantom{ }\\\Large \texttt{MMSSL}*},
    tick align=outside,
    tick pos=left,
    x grid style={darkgray176},
    xmin=6, xmax=94,
    ymin=0.075, ymax=0.1,
    xtick style={color=black},
    xticklabels={10\%, 20\%, 30\%, 40\%, 50\%, 60\%, 70\%, 80\%, 90\%},
    xtick={10, 20, 30, 40, 50, 60, 70, 80, 90},
    ytick={0.08, 0.085, 0.09, 0.095},
    y grid style={darkgray176},
    ytick style={color=black}
]


\addplot [thick, blue, forget plot, mark=*]
table {%
10 0.09672928
20 0.09603995
30 0.0943852
40 0.09169269
50 0.08904679
60 0.08775296
70 0.08472325
80 0.08441244
90 0.08360144
};


\addplot [thick, red, forget plot, mark=*]
table {%
10 0.09692852
20 0.09477874
30 0.09483766
40 0.09149738
50 0.08887534
60 0.0860732
70 0.08394822
80 0.08434773
90 0.08124682
};


\addplot [thick, green, forget plot, mark=*]
table {%
10 0.0951863
20 0.09413081
30 0.09334132
40 0.09131009
50 0.08857732
60 0.08907937
70 0.08745146
80 0.08534525
90 0.08182661
};


\addplot [thick, orange, forget plot, mark=*]
table {%
10 0.09811294
20 0.09696897
30 0.09661733
40 0.0951135
50 0.09462121
60 0.09409991
70 0.09220639
80 0.09014293
90 0.08775643
};

\end{axis}

\end{tikzpicture} \quad
    \begin{tikzpicture}

\begin{axis}[
    yticklabel style={
    /pgf/number format/.cd,
    fixed, fixed zerofill,
    precision=3
    },
    width=0.6\textwidth,
    height=0.3\textwidth,
    scaled y ticks=false,
    ylabel=Recall@20,
    xlabel=\makecell{Missing items (\%)\\\phantom{ }\\\Large \texttt{FREEDOM}},
    tick align=outside,
    tick pos=left,
    x grid style={darkgray176},
    xmin=6, xmax=94,
    ymin=0.075, ymax=0.1,
    xtick style={color=black},
    xticklabels={10\%, 20\%, 30\%, 40\%, 50\%, 60\%, 70\%, 80\%, 90\%},
    xtick={10, 20, 30, 40, 50, 60, 70, 80, 90},
    ytick={0.08, 0.085, 0.09, 0.095},
    y grid style={darkgray176},
    ytick style={color=black}
]


\path [draw=blue, thick]
(axis cs:10,0.0823979657264626)
--(axis cs:10,0.0836572328512873);

\path [draw=blue, thick]
(axis cs:20,0.082908023994662)
--(axis cs:20,0.0836132769924441);

\path [draw=blue, thick]
(axis cs:30,0.0816670755217469)
--(axis cs:30,0.08262513043478);

\path [draw=blue, thick]
(axis cs:40,0.0818646895122195)
--(axis cs:40,0.0834935130275453);

\path [draw=blue, thick]
(axis cs:50,0.0817109697411787)
--(axis cs:50,0.0838743168152549);

\path [draw=blue, thick]
(axis cs:60,0.082186140786587)
--(axis cs:60,0.0832041968688009);

\path [draw=blue, thick]
(axis cs:70,0.0831035433741179)
--(axis cs:70,0.0840757828075114);

\path [draw=blue, thick]
(axis cs:80,0.0827018941412073)
--(axis cs:80,0.0843566152498464);

\path [draw=blue, thick]
(axis cs:90,0.0835930203385619)
--(axis cs:90,0.0840606133674677);

\addplot [thick, blue, forget plot, mark=*]
table {%
10 0.083027599288875
20 0.0832606504935531
30 0.0821461029782634
40 0.0826791012698824
50 0.0827926432782168
60 0.0826951688276939
70 0.0835896630908146
80 0.0835292546955268
90 0.0838268168530148
};


\path [draw=red, thick]
(axis cs:10,0.0945630266259568)
--(axis cs:10,0.0961981981341397);

\path [draw=red, thick]
(axis cs:20,0.0919869305070267)
--(axis cs:20,0.0939714664078612);

\path [draw=red, thick]
(axis cs:30,0.0896475388811622)
--(axis cs:30,0.0918087200666781);

\path [draw=red, thick]
(axis cs:40,0.08779301661834)
--(axis cs:40,0.0898419737567472);

\path [draw=red, thick]
(axis cs:50,0.0856149181832231)
--(axis cs:50,0.0873890007150723);

\path [draw=red, thick]
(axis cs:60,0.0847606272701017)
--(axis cs:60,0.0864973495831666);

\path [draw=red, thick]
(axis cs:70,0.0837195385930574)
--(axis cs:70,0.085861189857545);

\path [draw=red, thick]
(axis cs:80,0.0823988522836177)
--(axis cs:80,0.0840024841493865);

\path [draw=red, thick]
(axis cs:90,0.0828309726977447)
--(axis cs:90,0.0839110258525195);

\addplot [thick, red, forget plot, mark=*]
table {%
10 0.0953806123800482
20 0.0929791984574439
30 0.0907281294739202
40 0.0888174951875436
50 0.0865019594491477
60 0.0856289884266341
70 0.0847903642253012
80 0.0832006682165021
90 0.0833709992751321
};


\path [draw=green, thick]
(axis cs:10,0.0957137411178835)
--(axis cs:10,0.0969489385281219);

\path [draw=green, thick]
(axis cs:20,0.0928738823562218)
--(axis cs:20,0.0946048992794218);

\path [draw=green, thick]
(axis cs:30,0.0901508997834069)
--(axis cs:30,0.0929849280728774);

\path [draw=green, thick]
(axis cs:40,0.0886176772080554)
--(axis cs:40,0.090288233085831);

\path [draw=green, thick]
(axis cs:50,0.0841379961732711)
--(axis cs:50,0.0865431776679693);

\path [draw=green, thick]
(axis cs:60,0.0832204048257791)
--(axis cs:60,0.0845923186123042);

\path [draw=green, thick]
(axis cs:70,0.0812009225604156)
--(axis cs:70,0.0821316833296083);

\path [draw=green, thick]
(axis cs:80,0.0789938652367221)
--(axis cs:80,0.0810641278652221);

\path [draw=green, thick]
(axis cs:90,0.0781355831741958)
--(axis cs:90,0.080516676950139);

\addplot [thick, green, forget plot, mark=*]
table {%
10 0.0963313398230027
20 0.0937393908178218
30 0.0915679139281421
40 0.0894529551469432
50 0.0853405869206202
60 0.0839063617190417
70 0.0816663029450119
80 0.0800289965509721
90 0.0793261300621674
};


\path [draw=orange, thick]
(axis cs:10,0.0953973340704643)
--(axis cs:10,0.0962909761276545);

\path [draw=orange, thick]
(axis cs:20,0.0950776735882)
--(axis cs:20,0.096208172680402);

\path [draw=orange, thick]
(axis cs:30,0.0947307942015967)
--(axis cs:30,0.0955475784714056);

\path [draw=orange, thick]
(axis cs:40,0.09301591296758)
--(axis cs:40,0.094537947846149);

\path [draw=orange, thick]
(axis cs:50,0.0909331427133295)
--(axis cs:50,0.092333242333521);

\path [draw=orange, thick]
(axis cs:60,0.0911129526756233)
--(axis cs:60,0.0924691051425969);

\path [draw=orange, thick]
(axis cs:70,0.0889312957768376)
--(axis cs:70,0.0900578169474808);

\path [draw=orange, thick]
(axis cs:80,0.0856047506172517)
--(axis cs:80,0.0867743486555785);

\path [draw=orange, thick]
(axis cs:90,0.0833679460418521)
--(axis cs:90,0.0841676558300046);

\addplot [thick, orange, forget plot, mark=*]
table {%
10 0.0958441550990594
20 0.095642923134301
30 0.0951391863365011
40 0.0937769304068645
50 0.0916331925234252
60 0.0917910289091101
70 0.0894945563621592
80 0.0861895496364151
90 0.0837678009359284
};

\end{axis}

\end{tikzpicture}
}
\end{adjustbox}

\vspace{4mm}

\begin{adjustbox}{width=0.35\textwidth,center}
\begin{tikzpicture}
        \begin{customlegend}[legend columns=4,
        legend entries={Zeros, Mean, Random, FeatProp}]
        \addlegendimage{blue, mark size=3pt, mark=*}
        \addlegendimage{red, mark size=3pt, mark=*}
        \addlegendimage{green, mark size=3pt, mark=*}
        \addlegendimage{orange, mark size=3pt, mark=*}
        \end{customlegend}
    \end{tikzpicture}
\end{adjustbox}
\caption{Recommendation performance variation (calculated as Recall@20) in various settings of datasets, multimodal recommendation models, and missing modalities imputing strategies. Results of those models with * are not computed via mean and standard deviation over the whole set of 5 random samples due to the excessive computational time.}\label{fig:rq1}
\end{figure*}

\Cref{fig:rq1} displays the results of our investigation, comparing the performance of the FeatPop strategy to Zeros, Mean, and Random used on top of \texttt{MMSSL} and \texttt{FREEDOM}, for the Amazon Baby, Toys, and Sports datasets (see again~\Cref{sec:miss_modes}).

Overall, FeatProp provides superior performance to the other baselines on the vast majority of configurations. When employed with \texttt{MMSSL}, it consistently emerges as the optimal strategy for mitigating the lack of features across all missing percentages and datasets. Even though the same assertion is generally valid also for \texttt{FREEDOM}, the trend (in this case) is not consistent across missing rates and datasets. Indeed, there are specific scenarios with high missing items percentages (e.g., 80\% or 90\%) where FeatProp aligns or performs slightly worse than the other baselines (especially on Amazon Baby and Toys). However, FeatProp becomes again a winning solution on Amazon Sports, where it is outperformed by the other baselines only in the case of 90\% of missing items. 

Regarding the other baselines, the careful reader will note that Random yields (overall) the poorest contribution. Conversely, it is interesting to notice the unexpected behaviour of Zeros. Indeed, we observe that not only it ranks second across all combinations when missing item percentage exceeds 70\%, but it also provides a slight increase in the performance when applied to \texttt{FREEDOM} (i.e., +3.04\% on average for Amazon Baby and Toys from 10\% to 90\% of missing items). This comes as a surprise if we consider that the common (and obvious) trend involves a recommendation performance degradation with higher missing items percentages (e.g., see FeatProp on the \texttt{FREEDOM} model, whose performance lowers, on average, of -15.96\% from 10\% to 90\% of missing items).

\subsection{The impact of propagation layers}

\begin{table*}[!t]
    \caption{Ablation study results (measured as Recall@20) with varying numbers of propagation layers $L$ for the FeatProp algorithm on the Amazon Baby dataset. Boldface and \underline{underline} stand for best and second-to-best settings, respectively.}\label{tab:ablation}
    \centering
    \begin{adjustbox}
    {width=\textwidth,center}
    \begin{tabular}{llccccc}
    \toprule
    \multirow{2}{*}{\textbf{Model}} & \multirow{2}{*}{\textbf{Setting}} & \multicolumn{5}{c}{\textbf{Missing items (\%)}}\\ \cmidrule(lr){3-7}   
    & & 10\% & 30\% & 50\% & 70\% & 90\%\\ \cmidrule{1-7}
    \multirow{4}{*}{\texttt{MMSSL}} & $L = 1$ & \underline{0.0925} $\pm 6\cdot10^{-04}$ & 0.0905 $\pm 6\cdot10^{-04}$ & 0.0878 $\pm 8\cdot10^{-04}$ & 0.0845 $\pm 1\cdot10^{-03}$ & 0.0767 $\pm 1\cdot10^{-03}$\\
    & $L = 2$ & \textbf{0.0926} $\pm 6\cdot10^{-04}$ & 0.0906 $\pm 5\cdot10^{-04}$ & \underline{0.0882} $\pm 6\cdot10^{-04}$ & \underline{0.0853} $\pm 2\cdot10^{-03}$ & 0.0787 $\pm 9\cdot10^{-04}$\\
    & $L = 3$ & 0.0921 $\pm 1\cdot10^{-03}$ & \underline{0.0912} $\pm 6\cdot10^{-04}$ & \underline{0.0882} $\pm 8\cdot10^{-04}$ & \textbf{0.0856} $\pm 2\cdot10^{-03}$ & \underline{0.0789} $\pm 1\cdot10^{-03}$\\
    & $L = 20$ & 0.0921 $\pm 8\cdot10^{-04}$ & \textbf{0.0915} $\pm 8\cdot10^{-04}$ & \textbf{0.0883} $\pm 7\cdot10^{-04}$ & \underline{0.0853} $\pm 2\cdot10^{-03}$ & \textbf{0.0797} $\pm 1\cdot10^{-03}$\\
    \cmidrule{1-7}
    \multirow{4}{*}{\texttt{FREEDOM}} & $L = 1$ & \textbf{0.0855} $\pm 9\cdot10^{-04}$ & 0.0834 $\pm 1\cdot10^{-03}$ & \underline{0.0808} $\pm 1\cdot10^{-03}$ & 0.0771 $\pm 1\cdot10^{-03}$ & \textbf{0.0744} $\pm 7\cdot10^{-04}$\\
    & $L = 2$ & 0.0852 $\pm 9\cdot10^{-04}$ & \textbf{0.0842} $\pm 1\cdot10^{-03}$ & 0.0806 $\pm 1\cdot10^{-03}$ & \textbf{0.0783} $\pm 1\cdot10^{-03}$ & 0.0710 $\pm 6\cdot10^{-04}$\\
    & $L = 3$ & \textbf{0.0855} $\pm 1\cdot10^{-03}$ & 0.0839 $\pm 7\cdot10^{-04}$ & \textbf{0.0810} $\pm 1\cdot10^{-03}$ & \underline{0.0779} $\pm 6\cdot10^{-04}$ & 0.0710 $\pm 1\cdot10^{-03}$\\
    & $L = 20$ & \underline{0.0853} $\pm 1\cdot10^{-03}$ & \underline{0.0840} $\pm 1\cdot10^{-03}$ & 0.0805 $\pm 3\cdot10^{-04}$ & 0.0773 $\pm 6\cdot10^{-04}$ & \underline{0.0725} $\pm 7\cdot10^{-04}$\\
    \bottomrule
    \end{tabular}
    \end{adjustbox}
\end{table*}

In their work~\cite{DBLP:conf/log/RossiK0C0B22}, the authors assert that FeatPop can propagate feature embeddings for several propagation layers, as it effectively avoids oversmoothing~\cite{DBLP:conf/iclr/OonoS20} by upholding feature diversity through the boundary conditions of its diffusion equation. To validate this assertion also in our setting, we conduct experiments using varying values of propagation layers (i.e., $L \in [1, 2, 3, 20]$). The final results on Amazon Baby are displayed in~\Cref{tab:ablation}. From a careful analysis, a dual characterisation emerges. On the one hand, when employing the \texttt{MMSSL} model, the Recall@20 metric attains satisfactory levels through several propagation layers (e.g., 20 layers) especially with substantial proportions of missing items; this confirms the assumption from the original FeatProp work~\cite{DBLP:conf/log/RossiK0C0B22}. On the other hand, this trend is less evident with the \texttt{FREEDOM} model. The explanation for this duality lies in the intrinsic rationales and formalisations of the models under examination rather than the efficacy of our methodology. In this respect, it should be noticed that \texttt{FREEDOM} (by-design) already performs a multimodal propagation on the item-item graph. Intuitively, this means the FeatProp algorithm may require few propagation layers when applied to \texttt{FREEDOM} to reach convergence and eventually produce satisfactory recommendation performance.
\section{Summary}
This paper introduces, formalises, and addresses the problem of missing modalities in multimodal recommendation, a challenge that is poorly debated in the related literature. First, we formally define the problem setting, where we assume a percentage of items from the catalogue lacks all multimodal features. Second, inspired by the recent advances in graph representation learning, we propose to re-cast the missing modalities problem as a problem of missing graph node features, where established baselines exist. We provide a re-implementation of the original FeatProp algorithm which builds on the multimodal similarities between items in the projected item-item graph. An extensive experimental setting shows that, when placed on top of two recent multimodal recommendation systems, FeatProp can outperform the baselines, while a study on the number of propagation layers demonstrates that not all backbones may require the same propagation layers to converge. 

\begin{acks}
Daniele Malitesta and Fragkiskos D. Malliaros acknowledge the support by ANR (French National Research Agency) under the JCJC project GraphIA (ANR-20-CE23-0009-01). 
\end{acks}

\bibliographystyle{ACM-Reference-Format}
\bibliography{bibliography}

\end{document}